# A Case Study of the 2016 Korean Cyber Command Compromise

Kyoung Jae Park, Sung Mi Park, Joshua I. James
Legal Informatics and Forensic Science Institute
Hallym University, Chuncheon, South Korea
kyoungjae.park@hallym.ac.kr
2ricecake@gmail.com
joshua.i.james@hallym.ac.kr

**Abstract:** On October 2016 the South Korean cyber military unit was the victim of a successful cyber attack that allowed access to internal networks. Per usual with large scale attacks against South Korean entities, the hack was immediately attributed to North Korea. Also, per other large-scale cyber security incidents, the same types of 'evidence' were used for attribution purposes. Disclosed methods of attribution provide weak evidence, and the procedure Korean organizations tend to use for information disclosure lead many to question any conclusions. We will analyze and discuss a number of issues with the current way that South Korean organizations disclose cyber attack information to the public. A time line of events and disclosures will be constructed and analyzed in the context of appropriate measures for cyber warfare. Finally, we will examine the South Korean cyber military attack in terms previously proposed cyber warfare response guidelines. Specifically, whether any of the guidelines can be applied to this real-world case, and if so, is South Korea justified in declaring war based on the most recent cyber attack.

## 1. Introduction

On a theoretical level cybercrime and cyberwarfare are quite different. Each have different motivations and potential consequences. However, the methods used by both cyber criminals and government-sponsored attackers are - in most cases - nearly indistinguishable. Although there is currently much discussion about cyber warfare and appropriate or justified response, it is unclear how to apply even the most practical recommendations. This work gives a case example of a recent cyber attack against the South Korean military cyber unit, and analyzes Korea's response options in terms of cyber-conflict guides vs. reality.

South Korea has a 92.1% Internet penetration rate ("Asia Internet Usage Stats", 2016). Most banking and government services are offered online, and online shopping is extremely common. Likewise, credit cards and other virtual currencies are common throughout the country.

Korea's cyber security policy is largely based on isolationism and obscurity. Influence from such policies is seen in the well-meaning Digital Signature Act (1999). This Act mandated centralized cybersecurity practices for online banking and ecommerce on the provider and client sides, implemented in mandatory, government-controlled software. The policy continues to force the use of (no longer supported) browsers and browser plugins for conducting basic transactions online. Until a 2014 revision of the Act, alternative software using current cyber security best practices was prohibited. As of 2017, much online commerce and government online services in South Korea continue to require end-of-support software on the client side.

Although South Korea is seen as a highly-connected, technologically advanced country, policies such as these have stunted Korea's development in practical information security. The government is slow to change, especially after providing complete backing for an already implemented system. Further, the current system is extremely useful for law enforcement and intelligence services during investigations. Allowing a free market on information security practices would reduce direct access from the government.

These issues are not meant to be a criticism per se, but to give background context on the state of thinking behind cyber security decisions. Because of a national security focus on internal threats rather than external threats ("National Security Act", 2011), there is little motivation for security-focused change based on external pressures.

This situation is made worse because of alleged cyber conflicts between North and South Korea. There have been many cyber attacks against South Korean infrastructure in the past, which will be discussed later. Out of these attacks the South Korean government claimed that North Korea is behind all major incidents. However, evidence implicating North Korea is often circumstantial. Key evidence that allegedly proves North Korea's involvement is consistently classified as secret, and is never released to the public. This is similar to the FBI discussing evidence the Sony Pictures hacking case of 2014, also allegedly involving North Korea.

In this work, we examine major cyber attacks against South Korean organizations, specifically focusing on the 2016 attack on South Korea's cyber military unit; the 'Cyber Command'. The Korean Cyber Command revealed that their anti-virus relay servers were compromised on October 5th 2016, allegedly by North Korea. Confidential documents were exfiltrated during the attack (Kim, 2016).

## 2. Background
The Cyber Command was created under the Korean Defense Intelligence Agency (DIA) on January 1, 2010 in response to the 7.7DDOS (July 1, 2009) attack allegedly committed by North Korea. In September 2011, the Cyber Command came under the control of the Korean Ministry of National Defense which is in charge of the command and control of the Korean Armed Forces. Currently, there are an estimated 1,000 soldiers under the Cyber Command. The cyber command is divided into 4 primary units (publicly). There are:

- Research and Development (Corps 31)
- Cyber warfare (Corps 510)
- Psychological warfare (Corps 530)
- Education and Training (Corps 590)

The organization structure of the cyber command is shown in Figure 1.

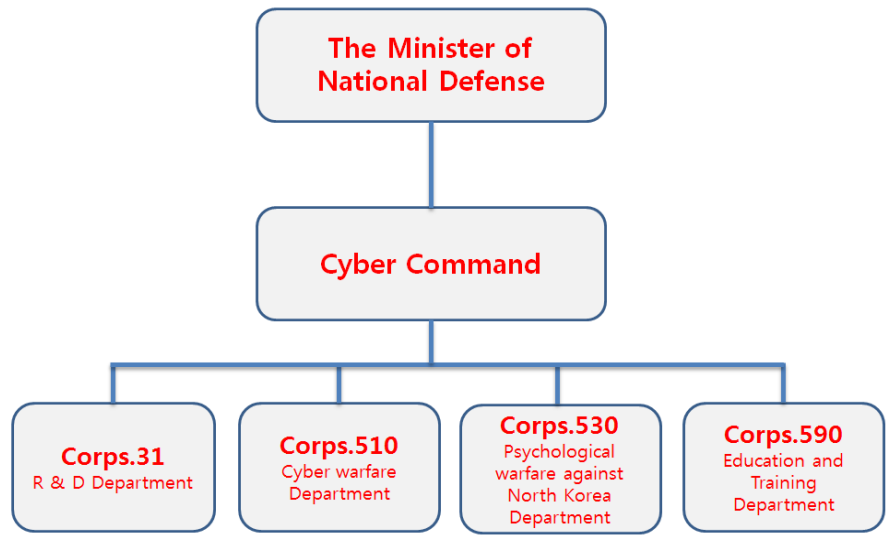

**Figure 1:** South Korean Cyber Command organizational structure. Stevenh. (2016) South Korean cyberwarfare unit for cyberwar. [사이버전(戰)을 위한 남북한 '사이버전 전담부대'] . Available at: http://ko.stevenh.wikidok.net/wp-d/585b33ef6926962302ff059a  [Accessed January 27, 2017].

**2.1 Major attacks against South Korea**
South Korean government agencies, businesses and critical infrastructure have been the victims of many large scale cyber attacks. On July 4, 2009, DDOS attacks were launched against major sites in the United States (Lee, 2009). On July 7, the attacks began to target major sites in South Korea. The attack took websites, including government websites, offline for up to 4 hours. The White House and major sites in the US, and South Korean websites of Media companies, political parties, the National Intelligence Service (NIS) and other major portal sites were targeted (Internet News Team, 2009). The NIS announced that it was most likely North Korean sponsored hacker groups. However, there was not enough direct evidence to conclusively link to North Korea. On October 30, 2009, during an audit with the National Assembly's National Defense Committee the NIS chief stated that the IP coordinating the attack was allocated to China. North Korea's telecommunication ministry was known to lease this IP from Chinese providers. Based on the evidence, North Korea was identified as the primary suspect in the 7.7 DDOS attack, a claim that North Korea denies (Kim, 2009).

"DarkSeoul" was a large-scale cyber terror attack that disabled the computer equipment of broadcasters and financial institutions in South Korea on March 20, 2013. Malicious code spread through the public to target critical systems through the compromise of popular Web servers. Popular sites were infected with drive-by malware (Kim, 2013). The malicious code took control of the PC or server inside the institution for up to 8 months. During this time the malware exfiltrated data, identified more vulnerabilities, and distributed malicious code. On March 20th, the malware deleted data in infected PCs and servers, all at the same time. At the time, more than 48,000 PCs were disabled. This resulted in media coverage and banking outages for ten days. As well as financial damage of an estimated 900 billion Korean Won (approx. 900 million USD) (Special Coverage Team, 2013). The South Korean government was also quick to name North Korea as the prime suspect. There are four main pieces of evidence that attributes the attack to North Korea. Sung and Lee (2013) describe them as:

1. Traces of manipulating an infected PCs through the same Chinese IP address used in previous attacks
2. Traces of a Korean-language PC accessing the financial companies 1590 times, distributing malicious code and leaking data since June 28, 2012
3. Malicious code similar to the previous hacking in South Korea where North Korea was suspected
4. An attack route that was the same as previous hacking that was allegedly by North Korea

Finally, on December 15, 2014, personal information of Korea Hydroelectric Nuclear Power Corporation (KHNP) staff members and confidential documents related to nuclear power plants and nuclear power plants (including drawings of nuclear power plants) were revealed on the hacker's' blog. The document contained the phrase "Who Am I?", and stated that they were an anti-nuclear group (Lee, 2014). They also announced a second attack during Christmas. The media claimed that the malicious code was similar to prior cyber terror attack, and that the attackers used expressions mainly used in North Korea, notifying the attack through Twitter (Jin, 2014). The hacker disclosed internal documents related to the nuclear power plant on December 15, 18, 19, 21, and 24, and requested to shut down Korea nuclear power plants Kori 1 and 3 and Wolsong 2 starting from Christmas. The attackers threatened to reveal 10 million pages of unpublished data and carry out the second attack if their demands were not met. On October 24, the South Korean government announced that the KHNP' attack originated in Shenyang, China. On March 17, 2015, Hanshin announced that the KHNP hacking attack was from a North Korean hacker organization (Koo, 2015). The evidence stated was:

1. The malicious code is similar to the "kimsuky" malicious code allegedly used by a North Korean hacker organization in a prior attack (not conclusively proven)
2. The same Chinese IP address used previously that is believed to also be used by North Korea

**2.2 Cyber Attack Against South Korean Cyber Military**
After establishing the Defense Integrated Data Center (DIDC) in 2014, a private company that built the computer network connected the internal computer system to the external computer system in the process of building the computer network. They finished the work without disconnecting. Through this route, malicious code penetrated the military network for the first time on August 4, 2016. On September 23, Cyber Command's vaccine routing server detected signs of malicious code infection, and the army separated the network of vaccine routing servers on September 25, midnight, and set up a joint investigation team on October 2 to investigate the infiltration route, the damage situation, and the identity of the hacker (Lee, 2016).

On December 5, a two-month investigation found that there was an intrusion in the intranet (defense network), and the possibility confidential documents were leaked. It was initially claimed that the first signs of infiltration were found on the Cyber Command's vaccine routing server, however it was revealed that on December 7, that traces of intrusion were also found on the vaccine routing server in Gye-Ryong-Dae DIDC (Defense Integrated Data Center). DIDC is the center for the information systems of the Army, Navy and Air Force.

The center has an important role, and the total number of infected computers was 3,200. 2,500 systems were external networks (Internet use), 700 were internal networks

(defense networks), and the defense minister's business computer was also infected with malware. The military revealed that the hacker's IP address was allocated to Shenyang, China; again suspected of being used by the North Korean military. Malware code similar to prior cases were also found (North Korea suspected). Also, a Hangul (the Korean alphabet) keyboard was used for compiling the malicious code. With this evidence, the military announced that North Korea was the likely suspect (Lee, 2016). On December 9, North Korea denied the claim and argued, "자신들이 해킹한 것이라면 왜 계속 같은 주소대역대의 IP 주소를 사용하겠냐" which translates as: *If we hacked them, why would we continue to use the IP address range*? According to North Korea, blaming them was an attempt to manipulate public opinion to divert from political issues in South Korea (Kim, 2016). On December 12, the Ministry of National Defense announced that the Defense Commission had leaked confidential data but could not disclose what kind of data it contained, and could only reveal that it was not a serious document (Lee and Ryu, 2016). However, a day later, on December 13, the military prosecutors and Defense Security Command proceeded to carry out search and seizure. It was recently revealed that confidential information was leaked. The investigation is still ongoing (Hong, 2016).

## 3. Appropriate Response from South Korea

The attack against the South Korea Cyber Command is an excellent case study to test out theoretical recommendations for cyber warfare response. The attack was targeted against a nation's military, successfully leaked confidential military documents and evidence leads back to only a few main suspects (though none proven beyond reasonable doubt). There is likely no cyber attack that has greater standing as a cyber conflict, beyond two nations openly admitting their actions. We will analyze this case in the context of the Tallinn Manual.

After North Korea was identified to be the suspect behind several cyber attacks targeting critical infrastructure, the people of South Korea started to consider cyber attacks as an "act of war" and asked for "retribution" in both cyber and traditional forms (Yoo, 2013). However, the issue is not as simple as people believe it to be. Not only is there the politics between countries to consider, but also the legal perspective. Retribution or declaring war is a complex, often prohibited, procedure. Meanwhile, it is also reasonable that a national crisis such as a severe cyber attack cannot continue to happen without any deterrence. That is why the Tallinn Manual garnered attention with the hopes of having a strict guideline in the new domain of cyber warfare

In this section, we will attempt to establish which rules of the Tallinn Manual could potentially apply to cyber attacks similar to South Korea's case. This section will be divided into the terms in which cyber attacks could be categorized, and what the victim State would be entitled to do accordingly.

### 3.1 Cyber operation as armed attack

The first Rule we need to consider is when the cyber operation is an armed attack within the meaning of the UN charter. Rule 13 of the Tallinn Manual (Schmitt, 2013) attempts to identify situations of legitimate self-defense. The right to exercise self-defense is not only limited to kinetic attacks, but also can be justified in case of cyber attacks (Rule 13, cmt. 3). Because of the term "armed" attack, however, the experts had varying opinions; it was disputed if weapons were necessary for an "armed" attack (Rule 13, cmt. 4). The majority of experts denied the necessity and were more inclined to set an analogy as to determine the severity of the cyber attack (Rule 13, cmt. 4). For them, a cyber operation could be considered as an armed attack if the consequences were equal to those of an attack with a kinetic weapon (Rule 13, cmt. 4). The idea that cyber operations could be sufficiently grave to qualify as armed attacks came to an accord unanimously (Rule 13, cmt. 3). The Rule itself defines armed attack as cyber operation depending on scale and effects. Based on the Nicaragua judgment (Nicaragua v. United States of America 1986, para. 191), the scales and effect are criteria to determine the "most grave forms" of use of force (Rule 13, cmt. 6). To be more detailed, according to the International Group of Experts, any use of force that "injures or kills persons or damages or destroys property" (Rule 13, cmt. 6) would be enough to constitute as armed attack. A more precise definition of what exactly would be "grave" is not given.

In the recent case of South Korea, this Rule cannot apply as comment 6 states "acts of cyber intelligence gathering and cyber theft, cyber operations that involve brief or periodic interruption of non-essential cyber services" are not considered to be armed attacks (Rule 13, cmt. 6).

### 3.2 Cyber operation as use of force

While use of force takes up an entire chapter of the Tallinn Manual, it is important to differentiate the terms "use of force" and "armed attack". These two standards are to be distinguished as they serve different purposes (Rule 11, cmt. 11). "Use of force" is defined with the simple goal to decide whether a State has violated the prohibition of the use of force according to Article 2(4) of the UN Charter (United Nations 1945) and related customary international laws (Rule 11, cmt. 11). "Armed attack", on the other hand, gives a State the right to defend itself with use of force (Rule 11, cmt. 11). Thus, any cyber operation that does not constitute as an armed attack, the victim State must resort to other methods, such as countermeasures in Rule 9 (Rule 11, cmt. 11). Responding in use of force without receiving an armed attack would violate the prohibition.

Rule 11 again uses the phrase "scale and effects", without further explanation. The comments make it clear the physical effects serve as a major criteria; States have denied inclusion of economic or political coercion as a use of force, so mere funding of a hacktivist group would not qualify (Rule 11, cmt. 3). Giving an organized group malware and training to use it to conduct cyber attacks against other States would be considered use of force (Rule 11, cmt. 4). Despite this, dealing with cyber operations without a kinetic equivalent is still unsettled. The experts provided a probable assessment with 8 factors to consider. This, however, is also not definitive; a holistic assessment of the operation depending on the various circumstances seems to be unavoidable (p.20, Schmitt, 2012).

### 3.3 State responsibility

This is a Rule that determines the scope of responsibility a State has for its cyber operation. Based on the customary international law of State responsibility, the State has international legal responsibility if the act is attributable to it and the act constitutes as a breach of an international obligation (Rule 6, cmt.2). Any cyber operations conducted by organs of a State, which is understood as a broad term, including every person of entity under the State's internal legislation, can be attributed to the State (Rule 6, cmt.6). If the organ appears to operate in official capacity and breach international obligations, the State becomes responsible (Rule 6, cmt.7). Whether the organ acted "in compliance with, beyond, or without any instructions" has little consequence (Rule 6, cmt.7). This is the same for other entities empowered by domestic law to be equivalent to governmental authority (Rule 6, cmt.8). However, standards to determine how much evidence "attributable" needs is unsettled. It stands to question if suspecting a State can be enough or if it is even possible to attribute the questionable act to a State with absolute certainty. In comment 14, acts can be retroactively attributed to the State if it acknowledges or adopt the conduct. It is unlikely in case of the recent attack of South Korea, that the North Korean government would admit to any involvement.

Internationally a wrongful act can refer to cyber operations that are violating international law (Rule 6, cmt.4). In case of cyber espionage, however, this becomes controversial as there is no international law prohibiting the act per se (Rule 6, cmt.4). One possible interpretation could be based on Rule 1 comment 6; cyber operation directed to another State's cyber infrastructure may violate the latter's sovereignty. However, whether malware is used for monitoring purposes only without causing physical damage could be regarded as breach of sovereignty was disputed without consensus (Rule 1, cmt. 6). It would depend on the severity and effects of the act to determine if cyber espionage is an international wrongful act at all.

Even if it is considered as an international wrongful act thereby justifying countermeasures in context of Rule 9, the appropriate responses are not at full consensus. It is the widespread agreement that cyber countermeasures cannot include the threat or use of force (Rule 11), however there are some who accept limited degree of military force. In comment 7, the Experts foresee a "proportionate" countermeasure to be appropriate, which is determined by the gravity of the initial unlawful act. If the origin of the cyber breach is unclear, the State may invoke a plea of necessity (Rule 9 cmt. 12). The State may be entitled to counter hacking, if the action is the "only way" to prevent further attacks and does not violate interests of other States too severely (Rule 9 cmt. 12). The precise limits and scope of this plea is still debated.

### 3.4 Conclusion
The Tallinn Manual, despite giving a guideline, does not give sufficient material in certain situations as stated in the scope section of the Introduction. However, even in dealing with cyber operations addressed in the manual, it does not seem to give enough criteria to determine the appropriate response if the cyber operation does not accompany physical consequences. Overall it can be concluded that terms used as standards are too ambiguous to avoid subjective assessments.

One thing that needs to be mentioned specific to South Korea's situation with North Korea, is that applying international law itself could be problematic. North Korea is a state with limited recognition; South Korea considers North Korea to be a part of the same nation and vice versa (Scofield, 2005). Therefore, in theory, should an offensive cyber operation be proven to be North Korea's doing, they would fall under Korean Criminal Law. However, in reality attempts to push South Korea's criminal law to North Korea would more likely be detrimental. Potential solutions and responses should be found within international law; even in cases without physical damages.

### 4. South Korea Response
As shown in prior cases, the evidence found in major cyber attacks are weakly linked to North Korea. While North Korea is certainly a prime suspect, many alternative hypotheses are just as likely. These alternatives, however, are not politically beneficial for South Korea. A major concern about the South Korean situation is how predictable the initial response is, combined with no follow-through to any real conclusion. For example, in most attacks North Korea was *confirmed* by the government to be the attacker within a week of the attack. Police and other investigations into the attack concluded weeks later, finding similar evidence to prior attacks. Related attack 'traces' are now publicly available, and somewhat easily forged.

This situation presents a number of response challenges. First, other entities, such as governments or organized crime could attack South Korea and attempt to make traces similar to those presumed to be North Korean. Second, successful cyber attacks against South Korea critical infrastructure is quite common. The government's stance is that the perpetrator is known, but the attacks continue to be successful. This demonstrates a lack of ability or willingness to secure critical infrastructure and control a known adversary. Finally, a major point of the South Korean response regarding almost all major cyber attacks is little transparency. North Korea is blamed, the same weak evidence is given and there is no further elaboration on methods, motives or prevention. The response, at least publicly, appears to be nothing.

An analysis in the context of the Tallinn Manual appears to show why South Korea is restricted in its response options. There are very few legal options for a country in South Korea's position. Ultimately, the issue is proper attribution, which is rarely possible but is a reality in cyber warfare. Without this attribution being unrealistically certain, most recommendations about cyber warfare response cannot be applied.

## 5. Conclusions

This work gave an overview of major cyber attacks against South Korea. These attacks work normally blamed on North Korea using similar evidential traces that are circumstantial. One attack that is presumed to be attributed to North Korea is used as a basis for claiming another attack must definitely be North Korea. The most recent attack that appears to be an act of cyber warfare was against the South Korean Cyber Command. We analyzed South Korea's response options to this real attack in the context of the Tallinn Manual, and found that South Korea has no military recourse, even from a targeted attack that leaked confidential information, if that attack is no longer ongoing. This partially explains South Korea's lack of a substantive reaction during any of the major attacks that took place.

### Acknowledgements
This research was supported by Hallym University Research Fund, 2016 (HRF201603007).